\documentclass[12pt,a4paper]{article}
\usepackage[dvips]{graphicx}
\usepackage[dvips]{epsfig}
\usepackage{achemso}
\newcommand{\beq}{\begin{equation}}
\newcommand{\eeq}{\end{equation}}

\newcommand{\dgr}{ {\,}^{\circ} \mbox{C}}
\newcommand{\form}{\mbox{C$_{12}$EO$_{5} \,$}}
\newcommand{\formeau}{\mbox{C$_{12}$EO$_{5}$/H$_{2}$O$\,$}}
\newcommand{\un}[1]{\ensuremath{\unskip\,\mathrm{#1}}}

\begin{document}

\title{\vspace*{-3cm}Magnetic nanorods confined in a lamellar lyotropic phase}
\author{Keevin B\'{e}neut, Doru Constantin\thanks{Author for correspondence. Email address:
constantin@lps.u-psud.fr},\\ Patrick Davidson and Arnaud Dessombz\\
Laboratoire de Physique des Solides, Univ. Paris-Sud,\\
 CNRS, UMR 8502, F-91405 Orsay Cedex, France.\\
 Corinne Chan\'{e}ac\\
 Laboratoire de Chimie de la Mati\`{e}re Condens\'{e}e, Univ. Paris VI,\\
  CNRS, UMR 7574, F-75252, Paris Cedex 05, France.\\}
\date{\today}

\maketitle

\begin{abstract}
The dilute lamellar phase of the nonionic surfactant \form was doped
with goethite (iron oxide) nanorods up to a fraction of 5~vol\%. The
interaction between the inclusions and the host phase was studied by
polarized optical microscopy (with or without an applied magnetic
field) and by small-angle x-ray scattering. We find that when the
orientation of the nanorods is modified using the magnetic field,
the texture of the lamellar phase changes accordingly; one can thus
induce a homeotropic--planar reorientation transition. On the other
hand, the lamellar phase induces an attractive interaction between
the nanorods. In more concentrated lamellar phases (under stronger
confinement) the particles form aggregates. This behaviour is not
encountered for a similar system doped with spherical particles,
emphasizing the role of particle shape in the interaction between
doping particles and the host phase.

\bigskip

\centerline{\epsfig{file=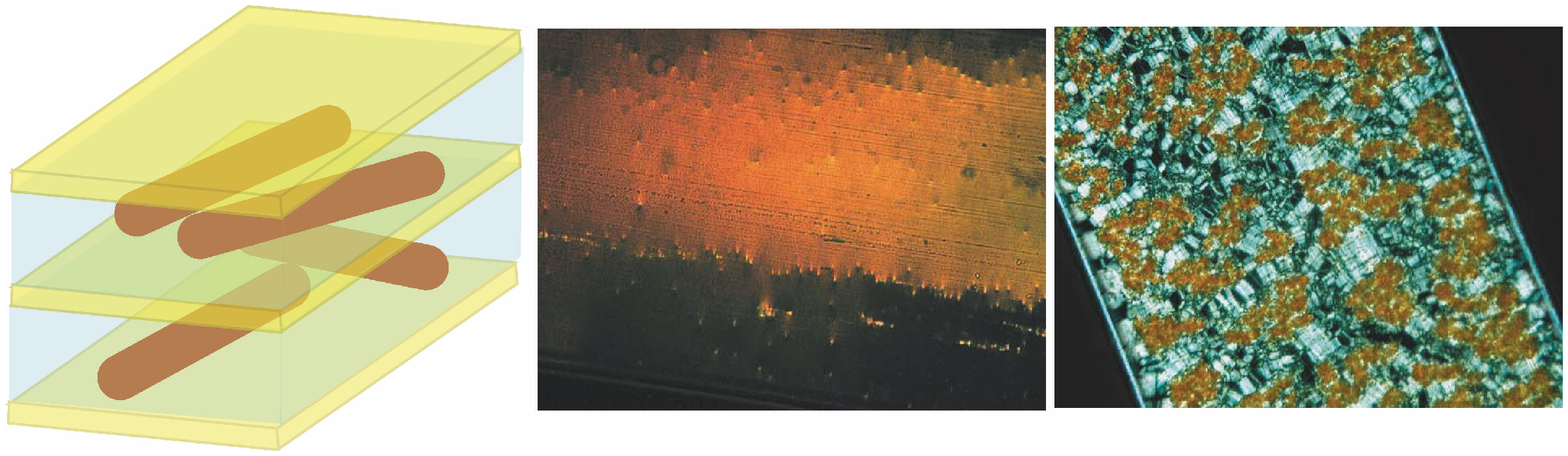,width=3.25in}}

\end{abstract}

\newpage

\section{Introduction}

In recent years, the field of soft matter physics has witnessed a
surge of activity in the area of hybrid organic--inorganic materials
\cite{Sanchez01}. This sustained interest was of course motivated by
the manifold applications of these systems \cite{MacLachlan00}, and
also by novel fundamental issues related to the interaction between
the two components. In many cases, the materials are obtained by
dispersing solid nanoparticles in a ``soft'' continuous matrix,
formed by surfactants, polymers, emulsions etc. By a careful choice
of the components, one tries to combine the specific properties of
the particles (catalytic, optical, magnetic etc.) and the
processability of the host phase.

Obviously, such dispersions also represent a new field among soft
matter systems; its novelty with respect to the ``classical''
colloidal solutions resides in the complexity of the matrix. Beyond
the theoretical interest of this study there is a very practical
one: for what parameter values (particle size and shape, structure
and elastic moduli of the matrix etc.) is the dispersion stable? Can
the confinement imposed by the host phase lead to ordering of the
inclusions? What are the ensuing applications? None of these
questions can be answered without a thorough understanding of the
interaction between the host phase and the inclusions.

In the case of nanoparticles dispersed in a liquid crystalline
matrix, one should naturally consider the effect of the elastic and
anisotropic medium on the interparticle potential. For lamellar
phases, the effect of smectic elasticity was modeled in detail
\cite{Turner97,Sens01,Evans03}, but the experimental studies are
still unsubstantial. Conversely, the inclusions can change the
interaction potential between membranes, and thus its repeat
distance \cite{Taulier00} and elastic moduli \cite{Ponsinet96}.
Clearly, in a composite system one must study:
\begin{itemize}
\item{}The influence of the confinement (due to the host phase) on the
inclusions.
\item{}The changes induced by the particles in the structure of the matrix.
\end{itemize}

Both these aspects are illustrated in a system that we formulated
recently: the host lamellar phase is the
C$_{12}$E$_{5}$/hexanol/water system, with C$_{12}$E$_{5}$ the
nonionic surfactant penta(ethylene glycol) monododecyl ether, and
the inclusions are iron oxide nanorods (goethite) with complex
magnetic properties \cite{Lemaire02}. We demonstrate the attractive
interaction between the particles induced by the lamellar matrix and
show that, when the confinement becomes too strong (i.e. the
lamellar repeat distance is too small) the particles aggregate, even
when their concentration is very low. On the other hand, we show the
action of the inclusions on the texture of the lamellar phase: when
the nanorods are oriented using an applied magnetic field, the
bilayers `follow' and (at high field) they align perpendicular to
the field. Perfect planar monodomains can thus be obtained, and the
alignment persists after removal of the field. The magnetic field
has no effect on the pure lamellar phase (without inclusions).

Lamellar lyotropic phases doped with small and spherical magnetic
particles have already been formulated \cite{Fabre90,Dabadie90} and
their structural \cite{Ponsinet96,Ramos96} and magnetic
\cite{Ponsinet95,Spoliansky00} properties were studied in detail
long ago. In contrast, our study deals with large, anisotropic
particles and with their interaction due to the confinement in the
lamellar phase. It is also noteworthy that gold nanorods can be
confined in lamellar phases of block copolymers, as reported
recently \cite{Deshmukh07}.

This hybrid system appears promising for the preparation of surface
layers of magnetic nanoparticles with well-defined spacing and
orientation (controlled by the host-induced interaction), with
applications, for instance, in high-density storage
media\cite{Gider96,Sun02}. The viscoelastic properties of the
lamellar phase are also interesting in view of fine-tuning the
deposition process (e.g., by spin-coating).

\section{Experimental}

Goethite ($\alpha-\un{FeOOH}$) is an iron oxyhydroxide, widely used
as a pigment. The nanorods were synthesized according to
well-established protocols \cite{Atkinson67,Jolivet04}. Their
dimensions are of the order of $150 \times 25 \times 10 \,
\mathrm{nm} ^3\, $ (length $\times$ width $\times$ height)
\cite{Lemaire04b}. The surface of the particles is hydroxylated,
with a surface charge of $0.2 \un{C \, m^{-2}}$ at $pH = 3$ and with
an isoelectric point corresponding to $pH = 9$; see
\cite{Lemaire04b} for more details. Bulk goethite is
antiferromagnetic \cite{Coey95}, but the nanorods bear a permanent
magnetic dipole $\mu \sim 1200 \, \mu _{B}$ along their long axis,
probably due to uncompensated surface dipoles ($\mu _{B} = 9.274 \,
10^{-24}\, \un{J/T}$ is the Bohr magneton). Furthermore, the easy
magnetization axis is perpendicular to this direction so that, at
high applied fields, the induced magnetic moment overtakes the
permanent one and the orientation of the rods switches from parallel
to perpendicular to the field at a critical value $B \sim 250
\un{mT}$.

The surfactant, \form, was acquired from Nikko and the hexanol from
Fluka; they were used without further purification. The phase
diagram of the \formeau mixture was determined more than 20 years
ago \cite{Mitchell83}. Ever since, it was extensively studied due to
the presence of several mesophases, and especially of a lamellar
phase that can be swollen up to a few percent of membrane fraction.
This dilute lamellar phase appears at fairly high temperature, but
it was shown that it can be brought down to room temperature by the
addition of a co-surfactant such as hexanol
\cite{Freyssingeas96,Freyssingeas97}. We used a hexanol/\form ratio
of 0.35 by weight, corresponding to a molar ratio of 1.3 (hexanol
molecules for each surfactant molecule). The temperature domain of
the lamellar phase changes with dilution, but it extended at least
between 17 and $32 \dgr$ for all our samples. The membrane thickness
is $\delta \approx 2.9 \un{nm}$\cite{Freyssingeas96}.

Silica particles with a nominal diameter of 27~nm were obtained from
Sigma-Aldrich as concentrated colloidal suspensions (Ludox TMA 34)
in deionized water (34~wt.\%). We found a pH of 7 for the initial
suspension.

Concentrated stock solutions of
\mbox{C$_{12}$EO$_{5}$/hexanol/H$_{2}$O$\,$} were mixed with
colloidal suspensions (of goethite or silica) and deionized water to
yield the desired volume fractions of membranes and doping
particles. The samples were contained in flat glass capillaries,
50-100~$\mu$m thick (Vitrocom) and aligned in homeotropic anchoring
(by thermal annealing).

The magnetic field was applied using a home-made setup based on
permanent magnets with a variable gap. Fields of up to 0.9~T can be
achieved. For polarized microscopy observation we used an Olympus
BX51 microscope with a rotating stage and long working distance
objectives (with a $5\times$ or $10\times$ magnification.)

Small angle x-ray scattering (SAXS) experiments were performed at
the ID02 station of the European Synchrotron Radiation Facility
synchrotron in Grenoble, France. The incident beam had a wavelength
$\lambda = 0.0995\,\un{nm}$, and the sample--detector distance was
5~m. The scattered x-rays were detected with a specially developed
CCD camera. A detailed description of the experimental setup can be
found in reference \cite{Narayanan01}. The $q$ range over which the
data could be reliably collected was $0.018 < q < 0.6 \un{nm}^{-1}$.
The flat faces of the capillaries were set perpendicular to the
x-ray beam.

\section{Results and Discussion}

\subsection{Phase diagram of the doped system}

\begin{figure}[htbp]
\centerline{\epsfig{file=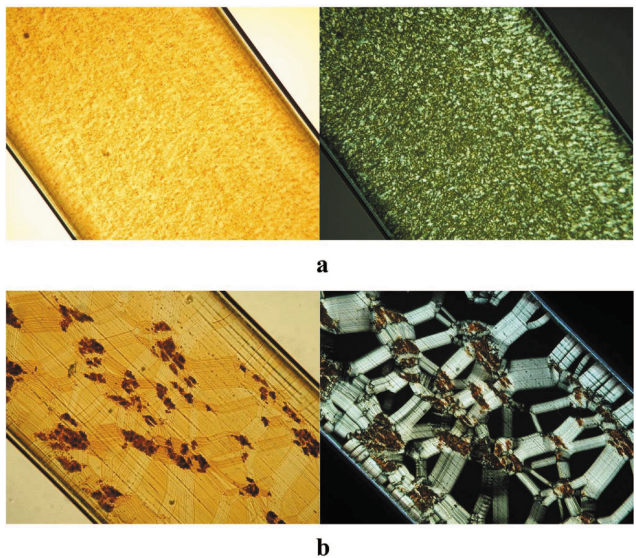,width=3.25in}}
\caption{\protect\small Lamellar phase doped with a goethite
concentration $\phi _g = 0.5$~\%, for a membrane concentration $\phi
_m = 7.2$~\% (a) and 10.2~\% (b), two weeks after preparation. Left:
in natural light. Right: between crossed polarizers. Aggregation of
the nanorods is observed in the more confined system (b).}
\label{fig:aggregation}
\end{figure}

The first step of the study was determining the phase diagram of the
system, more specifically the range of confinement (controlled by
the membrane volume fraction) for which the particles can be added
to the phase without demixing and their maximum concentration. We
started by preparing mixtures with a volume fraction of goethite
$\phi _g = 0.5$, 1, 1.5 and 2~\% and a membrane volume fraction
$\phi _m = 4.4$, 7.2, 10.2 and 14.4~\%. $\phi _g$ is defined as the
ratio between the volume of goethite particles and the total volume.
$\phi _m$ is the sum of the hexanol and surfactant volumes divided
by the total volume (we assume that mixing volumes are negligible).
The particles remained well dispersed in the dilute lamellar phase
($\phi _m = 4.4$ and 7.2~vol.~\%) for all values of $\phi _g$
investigated. In the concentrated phases, on the other hand,
particle aggregation was discernable after a few hours and was very
clear after a few days, even at the lowest particle concentration
(see Figure \ref{fig:aggregation}). We then prepared samples with
$\phi _m = 7.2$~vol.~\% and $\phi _g$ up to 5~vol.~\%. All these
samples have been stable for months \cite{endnote01}.

We conclude that a significant amount of goethite can be dispersed
in the lamellar phase as long as the membrane fraction remains below
a threshold in the interval $7.2 < \phi _m < 10.2$~vol~\%,
corresponding to a lamellar repeat distance $28 < d < 40 \un{nm}$.
The upper transition temperature of the lamellar phase (towards the
sponge phase) exhibits no significant variation as a function of the
doping fraction $\phi _g$. The mixing of the nanorods and the
lamellar phase presumably leads to an energy gain owing to the
formation of hydrogen bonds between the hydrated surface of the
particles and the surfactant heads \cite{Frost05}.

At this point, we have no convincing explanation for the threshold
value of the repeat distance. The most plausible connection is that,
as the lamellar phase becomes more concentrated, its elastic moduli
increase and so does the interaction between particles
\cite{Turner97}, to the point of inducing aggregation.

It should also be noted that the threshold value is of the order of
the particle width. An alternative explanation would therefore be
that aggregation occurs when rotation about the long axis of the
particles is hindered (and the particle loses a degree of freedom).
It is however not clear whether this explanation is compatible with
a strong interaction between the particles and the surfactant heads.

\subsection{Magnetic field effect}

\begin{figure*}[htbp]
\centerline{\epsfig{file=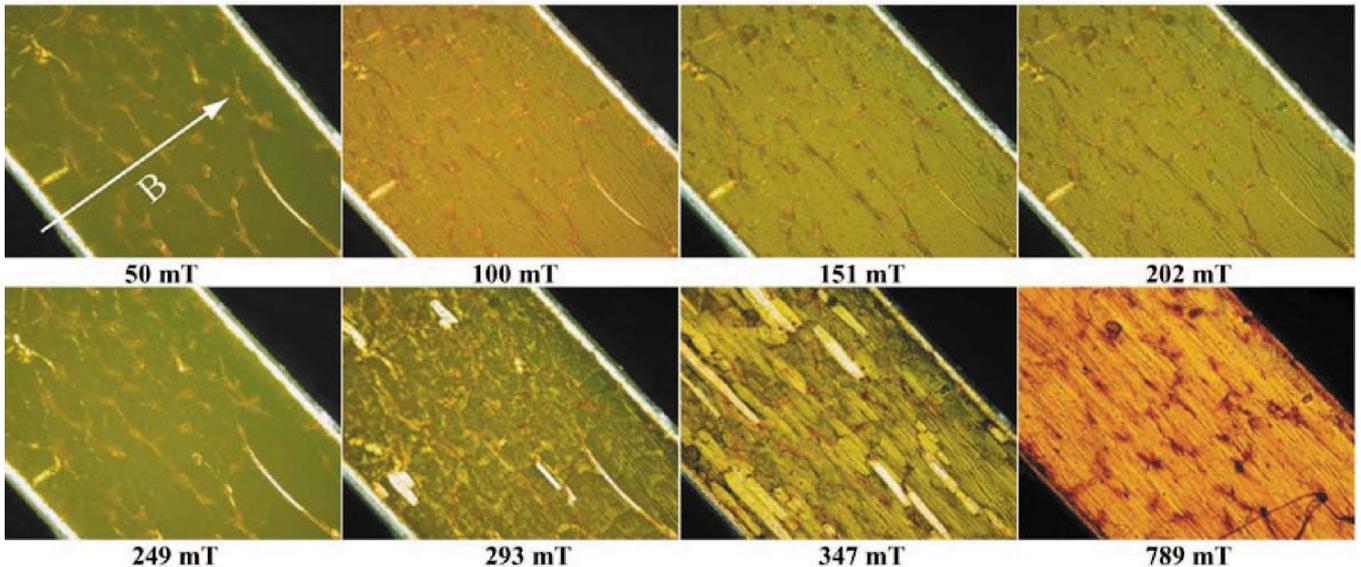,width=7in}}
\caption{\protect\small A magnetic field was applied across a
capillary of lamellar phase (membrane volume fraction 7.2 vol. \%)
doped with 1.5~vol.~\% nanorods. The field was increased from 0 to
about 0.8~T (the field direction is shown in the first image and its
values are given below each image). The images are taken between
crossed polarizers parallel to the sides of the
image.}\label{fig:field}
\end{figure*}

Magnetic field measurements were performed on flat glass
capillaries, 50~$\mu$m thick and 1~mm wide. The field was applied in
the plane of the capillary, perpendicular to its long axis.

We started by applying an increasing field (from 0 to about 0.8~T)
to a sample oriented in very good homeotropic anchoring (obtained by
annealing overnight close to the transition temperature to the
sponge phase). The membrane volume fraction was $\phi _m = 7.2$~\%
and the goethite volume fraction $\phi _g = 1.5$~\%. A few very thin
oily streaks persisted. The succession of images is shown in Figure
\ref{fig:field}. At low field, the transmitted intensity increased
with the field up to about 0.15~T (nanorods aligned along the
field); it then decreased to 0 at 0.25~T and increased again at
higher field, as the rods aligned perpendicular to the field.
Starting from the initial homeotropic anchoring, above 0.3~T the
existing oily streaks became more pronounced and new ones nucleated;
the texture gradually switched to planar anchoring, with the smectic
director along the field. This crossover corresponds to the value at
which the particle orientation changes from parallel to
perpendicular to the field in aqueous solution\cite{Lemaire02}.
Consequently, we infer that, as the particles turn, the lamellae
follow, presumably due to the strong association between the
goethite nanorods and the surfactant heads.

\begin{figure}[htbp]
\centerline{\epsfig{file=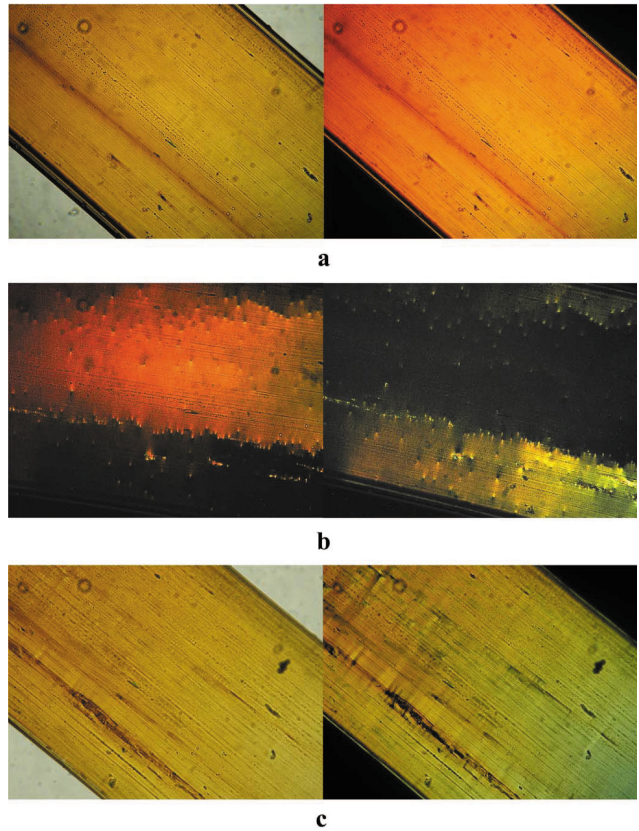,width=3.25in}}
\caption{\protect\small Two-domain area in a sample with $\phi _m =
4.4$~\% and $\phi _g = 1$~\%. a) Under strong field (834~mT), in
natural light (left) and between crossed polarizers parallel to the
sides of the image (right). b) By rotating the sample, total
extinction is obtained for each of the domains. c) After field
removal, the planar orientation persisted.} \label{fig:planar}
\end{figure}

Very good planar anchoring can thus be obtained, as shown in Figure
\ref{fig:planar}. The field was applied overnight; during this time,
the sample was kept at a temperature about $1 \dgr$ below the
transition temperature to the sponge phase, in an oven. The
temperature was then slowly decreased to its ambient value. Figure
\ref{fig:planar}a shows the sample, in natural light and between
crossed polarizers, under high field. It contains two domains
separated by a wall. By rotating the sample between the polarizers
(which remain parallel to the the sides of the photo) each domain
can be extinguished (Fig. \ref{fig:planar}b). The disorientation
between the domains can thus be estimated at $3.6 {\,}^{\circ}$.

The field was then progressively decreased (in steps of 0.1~T every
10 minutes). Some focal conic textures developed during the process,
but they annealed after a couple of hours. The resulting texture at
zero field is shown in Fig. \ref{fig:planar}c; it was stable for
days.

The effect of the magnetic field is similar to that observed in
ferrosmectic phases obtained by doping dilute lamellar phases with
small ferromagnetic particles \cite{Fabre90,Ponsinet95}. Indeed,
these authors also observe a reorientation transition, signaled by
the appearance of focal conic defects in homeotropic samples under
the influence of a magnetic field applied along their director axis
(normal to the layers). However, in their system the layers tend to
align {\textit along} the field, while in our case they prefer to be
{\textit perpendicular} to it above the critical value.

\subsection{Interaction induced by the lamellar phase}

We used x-ray scattering to study the interaction between colloidal
particles in the lamellar phase and in solution. It is
well-known\cite{Chaikin95} that the intensity scattered by a
collection of identical particles can usually be written as the
product $I(q) = | F(q) |^2 \times S(q)$ of a form factor, $| F(q)
|^2$, dependent only on the size and shape of the individual
particle, and a structure factor $S(q)$ quantifying the interactions
between particles ($S(q)=1$ in the absence of interactions).

\begin{figure}[htbp]
\centerline{\epsfig{file=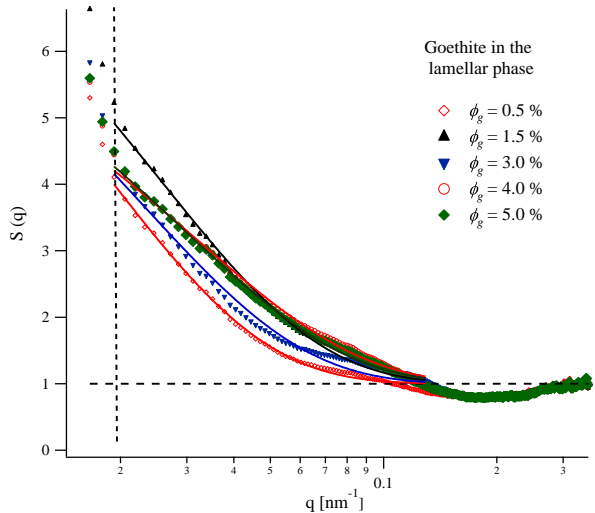,width=3.25in}}
\caption{\protect\small Structure factors for goethite particles
dispersed in the lamellar $L_{\alpha}$ phase. The membrane volume
fraction is $\phi _m = 7.2 \%$ and the volume fraction of goethite
$\phi _g$ is indicated for each curve. The vertical dashed line $q =
0.018\,\un{nm}^{-1}$ is the limit below which the data exhibits
marked uncertainty (due mainly to background subtraction). Solid
lines are exponential fits (see text).} \label{fig:goeth_lamellar}
\end{figure}

As form factor we used the intensity scattered by a very dilute
dispersion ($\phi _g = 0.066 \%$) in the lamellar phase at the same
membrane concentration as for the curves shown in Figure
\ref{fig:goeth_lamellar}, namely $\phi _m = 7.2 \%$. After
background subtraction, the scattering curves for ($L_{\alpha}\, +\,
$goethite) systems presented in the following were divided by this
signal and normalized to one at large $q$ vectors.

The first observation is that all structure factors shown in Figure
\ref{fig:goeth_lamellar} exhibit a marked increase at small angles
(below $0.1 \un{nm^{-1}}$), clear sign of a strong attractive
interaction. A quick estimate of the interaction range $\xi$ can be
obtained by fitting the data to an exponential decrease:
\begin{equation}
\label{eq:sq}
S(q) = 1 + A \exp (-q \xi) \, ,
\end{equation}
yielding $30 \un{nm} < \xi < 50 \un{nm}$. The attractive range is
similar to both the width of the nanorods and the lamellar repeat
distance. A more detailed study for different dilutions of the
lamellar phase is needed to assess the nature of the interaction.
The most interesting feature of this interaction is that it only
appears under confinement (in the lamellar phase), and for
anisotropic particles, as discussed below.

\begin{figure}[htbp]
\centerline{\epsfig{file=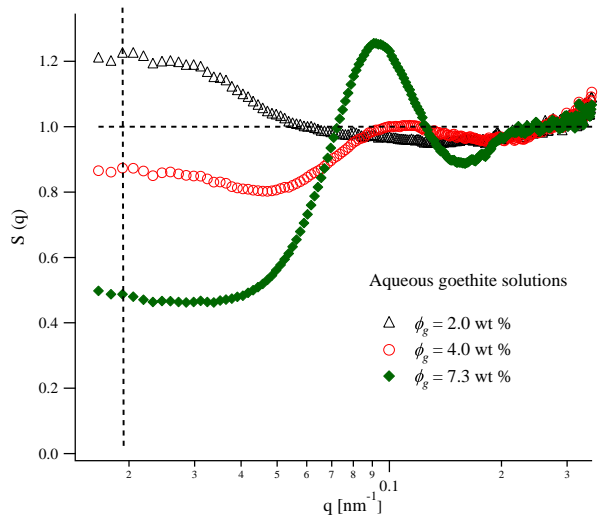,width=3.25in}}
\caption{\protect\small Structure factors for water dispersions of
goethite particles. The volume fraction of goethite $\phi _g$ is
indicated for each curve. The vertical dashed line $q =
0.018\,\un{nm}^{-1}$ is the limit below which the data exhibits
marked uncertainty (due mainly to background subtraction).}
\label{fig:goeth_water}
\end{figure}

As a reference system, we studied water dispersions of goethite
particles at similar concentrations, see Figure
\ref{fig:goeth_water}. The form factor used was obtained from a very
dilute aqueous solution, $\phi _g = 0.066 \%$. At lower particle
concentrations the structure factor is negligible; at $\phi _g = 7.3
\%$ a typical shape for hard-core systems, with a well-defined peak
and an oscillation at higher $q$ starts to appear, but its amplitude
is moderate and the shape very different from that measured in the
lamellar phase. We conclude that the effects described above (Fig.
\ref{fig:goeth_lamellar}) are due to the presence of the confining
lamellar phase.

The X-ray scattering data for the doped lamellar phases indicate
that the lamellar phase induces an attractive interaction between
the goethite nanorods. In order to assess the effect of shape we
studied the same lamellar phase doped with a comparable
concentration of silica beads, a system already described in the
literature \cite{Salamat99}. The membrane fraction was $\phi _m =
7$~vol\% and the volume fraction of beads was $\phi _s = 0.5$, 1, 2
and 3~vol\%. The samples were stable and homogeneous for months,
although in reference \cite{Salamat99}, the doped phase was found to
be stable only for volume fractions up to $\phi _s = 0.8$~vol\%.
This discrepancy might be related to the difference in the
presentation of the silica beads: they used Ludox TM solutions, at
pH~9 and with relatively high salt concentrations.

\begin{figure}[htbp]
\centerline{\epsfig{file=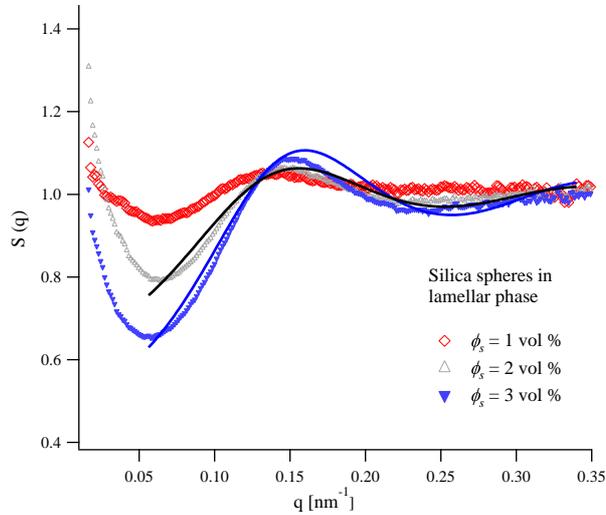,width=3.25in}}
\caption{\protect\small Structure factors for silica spheres
dispersed in the lamellar phase. The volume fraction of silica
particles $\phi _s$ is indicated for each curve. The data for $\phi
_s = 2$ and 3~vol\% is fitted with a Percus-Yevick hard sphere model
with an interaction radius of 19~nm.} \label{fig:silica_lamellar}
\end{figure}

Homeotropic doped samples were studied using the same procedure as
for the goethite-containing phases. The resulting structure factors
are shown in Figure \ref{fig:silica_lamellar}. The scattering
intensity $I(q)$ for the dilute sample ($\phi _s = 0.5$~vol\%) was
used as form factor; it is well described by a polydisperse sphere
model, with a radius $R \sim 13 \un{nm}$ and polydispersity
$p=\sigma / R \sim 0.1$. Similar values are obtained for the aqueous
dispersions using an in-house rotating anode setup (data not shown).

The first observation is that for the silica beads dispersed in the
lamellar phase the structure factors only exhibits a slight increase
at small angles. If present, the induced attraction is thus much
weaker for spheres than for rods. For wave vectors $q > 0.06
\un{nm}^{-1}$, the structure factors for $\phi _s = 2$ and 3~vol\%
are well described by a three-dimensional hard sphere interaction
(in the Percus-Yevick approximation \cite{Wertheim63,Thiele63}) with
an effective hard-core radius of 19~nm, see Figure
\ref{fig:silica_lamellar}. In conclusion, the presence of the
lamellar phase has no discernible effect on the interaction between
silica spheres. Rigorously speaking, for spheres confined between
rigid planes a two-dimensional (hard-disk) interaction would be a
more adequate description. We performed such an analysis using the
analytical form for the structure factor given by Y.
Rosenfeld\cite{Rosenfeld90} and obtain similar results, with a hard
disk radius of 17~nm. More concentrated systems would be needed to
discriminate between the 2D and 3D cases.

\section{Conclusion}

In conclusion, we formulated a nonionic lamellar phase doped with
large magnetic nanorods (in comparison with the interlayer
distance). The inclusions experience an attractive interaction under
confinement, a feature absent in simple aqueous solutions of similar
concentration or in systems of confined silica spheres. Under even
higher confinement (membrane concentration), the nanorods aggregate.
The interaction between the particles and the host phase is also
apparent in the orienting effect of the inclusions on the lamellar
phase.

\subsection*{Acknowledgements}

P. Panine, E. Belamie and A. Poulos are acknowledged for helping
with the synchrotron experiments.

\bibliographystyle{plain}
\begin{small}
\bibliography{ala_ref,goeth_ref}

\begin{mcitethebibliography}{31}
\bibitem[end()]{endnote01}
No change is visible in horizontally stored sealed capillaries. In vertically
  stored vials and capillaries, on the other hand, a concentration gradient
  appears, with a higher goethite concentration at the bottom. After briefly
  vortexing the vials, the solutions are again homogeneous, with no sign of
  aggregation.\relax
\mciteBstWouldAddEndPunctfalse
\mciteSetBstMidEndSepPunct{\mcitedefaultmidpunct}
{}{\mcitedefaultseppunct}\relax
\EndOfBibitem
\end{mcitethebibliography}
\end{small}


\end{document}